\documentclass[twocolumn]{article}
\usepackage{amsmath}
\usepackage{bm}
\usepackage{enumitem}
\usepackage{graphicx} 

\title{\textbf{The Application of Homotopy Perturbation Method to the Solution of Non-Linear Partial Differential Equations}}
\author{\LARGE\textbf{Gbenga Onifade Ebenezer}}
\date{Department of Mathematics, Faculty of Physical Sciences, University of Ilorin, P.M.B. 1515, Ilorin, Kwara State}

\begin{document}

\maketitle

\begin{center}
   {\LARGE\textbf{Abstract}}
\end{center}
In this study, a thorough investigation was conducted into the Homotopy Perturbation Method (HPM) and its application to solve the Burger and Blasius equations. The HPM is a mathematical technique that combines aspects of homotopy and perturbation methods. By introducing an auxiliary parameter, the complex problems were transformed into a series of simpler equations that could be solved step by step. The results of the study are significant. The HPM was found to be effective in solving the Burger and Blasius equations quickly and accurately. It proved to be a valuable tool for handling challenging mathematical problems. Consequently, researchers encountering seemingly impossible math puzzles can consider employing HPM as a potential solution.

\begin{center}
  \section{Introduction}  
\end{center}
In 1998, six years after Liao S.J. proposed the early 'Homotopy Analysis Method (HAM)' in his PhD dissertation, Jihuan He published the so-called 'Homotopy Perturbation Method (HPM).' Like the early HAM, HPM is based on constructing a homotopy equation.
\begin{multline}
(1-q)L[\Phi(x;q)-u_0]+qN[\phi(x;q)] = 0, \\
x\in\Omega, \quad q\in[0,1]
\end{multline}
Which is exactly the same as the zeroth order deformation equation. Like the HAM, the \(\phi(x;q)\) is also expanded into a Maclaurin Series.
\begin{equation}\Phi(x;q) = u_0(x) + \sum_{n=1}^{\infty} u_n(x)q^n \end{equation}\\
and approximation is achieved by setting \(q=1\), say,
\begin{equation} u(x) = u_0(x) + \sum_{n=1}^{\infty} u(x) \end{equation}\\
The only difference between HPM and the early HAM is that the embedding parameter \(q\in[0,1]\) is treated as a 'small parameter,' allowing the governing equation of \(u_n(x)\) to be obtained by substituting (2) into (1) and equating the coefficients of like powers of \(q\).\\
However, Hayat and Sayid proved in 2007 that, substituting the Maclaurin Series
\begin{equation}
    N[\phi(x;q)] = \sum_{n=0}^{\infty}D_nN[\phi(x;q)]q^n,
\end{equation}\\
where \(D_n = \frac{1}{2!}\frac{\partial^n}{\partial q^n}\) for \(q = 0\), and by substituting series (2) into (1) and then equating the coefficients of like powers of \(q\), one obtains:
\begin{equation}
L[u_n(x) - X_nu_{n-1}(x)] = D_{n-1}N[\phi(x;q)] 
\end{equation}
For \(u_n(x)\), which is exactly the same as the high-order deformation equation (5), no matter whether one expands the embedding parameter or not, one obtains the exact same approximations as the early HAM. Therefore, Sayid and Hayat pointed out that nothing is needed in Dr. He's approach except the new name HPM. Unfortunately, like the early HAM, the so-called HPM cannot guarantee the convergence of approximations, so it is valid only for weakly nonlinear problems with small physical parameters, as reported by many researchers. HPM has been used by Dr. He Ji Huan since he proposed the method in 1999 to solve:
\begin{enumerate}[label=\roman*.]
    \item Lighthill Equation
    \item Duffing Equation
    \item Non-Linear Wave Equation
    \item Schrödinger Equation
\end{enumerate}
In the perturbation technique, we will first propose a new perturbation technique coupled with the homotopy technique. In topology, two continuous functions from one topological space to another are called "homotopic." Formally, a homotopy between two continuous functions \(g\) and \(f\) from the topological space \(X\) to another topological space \(Y\) is defined to be a continuous function.

\begin{center}
$H : X \times [0,1] \longrightarrow Y$
\end{center}

This function is defined such that:

\begin{center}
$H(x,0) = f(x)$ and $H(x,1) = g(x)$ for all $x \in X$.
\end{center}

The HPM does not depend on a small parameter in the equation. In the homotopy technique in topology, a homotopy is constructed with an embedding parameter \(P \in [0,1]\), which is considered a small parameter.

\section{Basic Idea of Homotopy Perturbation Method}

Let us consider the nonlinear differential equation:

\begin{equation}
A(u) - F(r) = 0 \quad r \in \Omega
\end{equation}

with boundary conditions:

\begin{equation}
B\left(u, \frac{\partial u}{\partial \eta}\right) \quad r \in \Gamma
\end{equation}

Here, \(A\) is a general differential operator, and \(B\) is a boundary operator. \(\Gamma\) is the boundary domain of \(\Omega\), and \(F(r)\) is a known analytic function. The operator \(A\) can be divided into two parts, \(L\) and \(N\), where \(L\) is linear and \(N\) is nonlinear. Equation (6) can be written as follows:

\begin{equation}
L(u) + N(u) - F(r) = 0
\end{equation}

Using the homotopy technique, we construct a homotopy:

\begin{center}
$v(r,p) : \Omega \times [0,1] \longrightarrow R$
\end{center}

which satisfies:

\begin{multline}
H(v,p) = (1-p)[L(v)-L(u_0)] \\
+ p[A(v)-F(r)] = 0 \quad p \in [0,1]
\end{multline}

or

\begin{multline}
H(v,p) = L(v) - L(u_0) + pL(u_0) \\
+ p[N(v)-F(r)] = 0
\end{multline}
Where \(u_0\) is an initial approximation of Equation (6) that satisfies the boundary conditions. Obviously, from Equation (6), we will have:

\begin{equation}
    \text{for } p=0, \; H(v,0) = L(v) - L(u_0) = 0
\end{equation}

\begin{equation}
    \text{for } p=1, \; H(v,1) = A(v) - F(r) = 0
\end{equation}

The change in the value of \(p\) from zero to unity corresponds to that of \(v(r,p)\) from \(u_0(r)\) to \(u(r)\). In topology, this is referred to as deformation, and the terms \(L(v)-L(u_0)\) and \(A(v)-F(r)\) are called homotopic.

We will initially use the embedding parameter \(p\) as a small variable and assume that the solution of Equation (6) can be expressed as a power series of \(p\):

\begin{equation}
v = v_0 + pv_1 + pv_2 + \ldots
\end{equation}

Setting \(p\) to 1 results in the approximate solution of Equation (8):

\begin{equation}
u = \lim_{{p \to 1}} v = v_0 + v_1 + v_2 + \ldots
\end{equation}

The series (14) is generally convergent in most cases; however, the rate of convergence depends on the nonlinear operator \(A(v)\).
\section*{Application of Homotopy \hspace{0.5em}and Perturbation Method}
\subsection{Derivation of Blasius Equation}
For two dimensional, steady state, incompressible flow with a zero pressure gradient over a flat plate, the governing equations are simplified.
\begin{equation}
    \frac{\partial u}{\partial x} + \frac{\partial v}{\partial y} = 0
\end{equation}
\begin{equation}
    u\frac{\partial u}{\partial x} + v\frac{\partial v}{\partial y} = \frac{\partial^2 u}{\partial y^2}
\end{equation}
subject to boundary conditions:
\begin{center}
    $y = 0, \; u = 0$
\end{center}
\begin{equation}
    y \rightarrow \infty, \; u = u_\infty, \; \frac{\partial u}{\partial y} = 0
\end{equation}
To transform (15) and (16) into ordinary differential equations, take the stream function \(\phi\) defined by
\begin{equation}
    \phi = \sqrt{v_x u_\infty}f(\eta)
\end{equation}
where \(f\) is a dimensionless function of the similarity variable \(\eta\):
\begin{equation}
    \eta = \frac{y}{\sqrt{\frac{u_x}{u_\infty}}}
\end{equation}
Now,
\begin{center}
    $u = \frac{\partial \varphi}{\partial y} = \frac{\partial \varphi}{\partial \eta} \frac{\partial \eta}{\partial y}$
\end{center}
\begin{center}
    $ = \sqrt{v_x u_\infty}f'(\eta)\frac{1}{\sqrt{\frac{v_x}{u_\infty}}}$
\end{center}
\begin{equation}
    = u_\infty\frac{\partial f}{\partial \eta}
\end{equation}
Similarly,
\begin{center}
    $v = -\frac{\partial \varphi}{\partial x} = -\left[\frac{\partial \sqrt{v_x u_\infty}}{\partial x}f(\eta) + \sqrt{v_x u_\infty}\frac{\partial}{\partial x}f(\eta)\right]$
\end{center}

\begin{center}
    $= -\left[f(\eta)\frac{1}{2}\sqrt{\frac{v u_\infty}{x}} + \sqrt{v_x u_\infty}\frac{df}{d\eta}\left(\frac{-1}{2}\right)\frac{yx^{-\frac{3}{2}}}{\sqrt{\frac{v}{u_\infty}}}\right]$
\end{center}

\begin{center}
    $= -\left[\frac{1}{2}f(\eta)\sqrt{\frac{vu_\infty}{x}} - \frac{1}{2}\frac{u_\infty y}{x}\frac{df(\eta)}{d\eta}\right]$
\end{center}

\begin{equation}
= \frac{1}{2}\sqrt{\frac{vu_\infty}{x}}\left[\eta\frac{df}{d\eta} - f\right]
\end{equation}
Now,
\begin{center}
    $\frac{\partial u}{\partial x} = u_\infty\frac{d^2f}{d\eta^2}\frac{1}{\sqrt{\frac{v}{u_\infty}}}\left(\frac{1}{2}\right)^{-\frac{3}{2}}$
\end{center}
\begin{equation}
= -\frac{u_\infty}{2x}\eta\frac{d^2f}{d\eta^2}
\end{equation}
Where
\begin{center}
    $\frac{\partial u}{\partial x} = \frac{\partial}{\partial x}\left(\frac{\partial \varphi}{\partial y}\right) = \frac{\partial}{\partial \eta}\left(\frac{\partial \eta}{\partial y}\cdot\frac{\partial \varphi}{\partial \eta}\right)\cdot\frac{\partial \eta}{\partial x}$
\end{center}
and
\begin{center}
    $\frac{\partial u}{\partial y} = \frac{\partial}{\partial y}\left(\frac{\partial \varphi}{\partial y}\right) = \frac{\partial}{\partial \eta}\left(\frac{\partial \eta}{\partial y}\cdot\frac{\partial \varphi}{\partial \eta}\right)\cdot\frac{\partial \eta}{\partial y}$
\end{center}
\begin{equation}
\Longrightarrow \frac{\partial u}{\partial y} = u_\infty\frac{d^f}{d\eta^2}\cdot\frac{1}{\sqrt{\frac{v_x}{u_\infty}}}
\end{equation}
\begin{center}
    $\frac{\partial^2 u}{\partial y^2} = \frac{\partial}{\partial y}\left(u_\infty\sqrt{\frac{v_x}{u_\infty}}\cdot\frac{d^2f}{d\eta^2}\right)$
\end{center}

\begin{center}
    $\frac{\partial^2 u}{\partial y^2} = \frac{u_\infty}{\sqrt{\frac{v_x}{u_\infty}}}\left(\frac{d^3f}{d\eta^3}\cdot\frac{1}{\sqrt{\frac{v_x}{u_\infty}}}\right)$
\end{center}

\begin{equation}
\frac{\partial^2 u}{\partial y^2} = \frac{(u_\infty)^2}{v_x}\frac{d^3f}{d\eta^3}
\end{equation}
Putting these values in equation (16), we get
\begin{center}
    $u\frac{\partial u}{\partial x} + v\frac{\partial u}{\partial y} = \frac{\partial^2 u}{\partial y^2}$
\end{center}

\begin{center}
$u_\infty\frac{df}{d\eta}\left[-\frac{u_\infty}{2x}\eta\cdot\frac{d^2 f}{d\eta^2}\right] + \frac{1}{2}\sqrt{\frac{vu_x}{x}}\left[\eta\frac{df}{d\eta}-f\right]\cdot u_\infty\sqrt{\frac{v_x}{u_\infty}}\cdot\frac{d^2 f}{d\eta^2} = \frac{v(u_\infty)^2}{v_x}\frac{d^3 f}{d\eta^3}$
\end{center}

\begin{center}
$\Longrightarrow -\frac{(u_\infty)^2}{2x}\eta\frac{df}{d\eta}\cdot\frac{d^2f}{d\eta^2} + \frac{1}{2}\frac{u_\infty}{x}\left[\eta\frac{df}{d\eta}-f\right]\frac{d^2f}{d\eta^2} = \frac{u_\infty^2}{x}\cdot\frac{d^3f}{d\eta^3}$
\end{center}
\begin{equation}
    \frac{d^3f}{d\eta^3} + \frac{1}{2}f\cdot\frac{d^2f}{d\eta^2} = 0
\end{equation}
With boundary conditions:
\begin{center}
  $\eta = 0, \;\; f' = \frac{df}{d\eta} = 0$
\end{center}
\begin{equation}
    \eta \rightarrow \infty, \;\; \frac{df}{d\eta} = 0
\end{equation}

\subsection{Solution of Blasius Equation by Homotopy Perturbation Method}
So, to get a solution for equation (25) using the perturbation technique, we construct a homotopy
\begin{center}
    $v(r, p): \Omega \times [0, 1] \rightarrow {R}$
\end{center}
which satisfies
\begin{center}
    $H(v, p) = (1-p)[L(v) - L(u_0)] + P[A(v) - f(r)] = 0, \;\; p \in [0,1], \;\; r \in \Omega$
\end{center}
or
\begin{multline}
    H(v, p) = L(v) - L(u_0) + PL(u_0) + P[N(v) - f(r)] \\
    = 0
\end{multline}
where \(u_0\) is an initial approximation of equation (26) that satisfies the boundary condition. Now, from equation (25)
\begin{equation}
    (1-p)\left(\frac{\partial^3 f}{\partial \eta} - \frac{\partial^3 f}{\partial f_0}\right) + p\left(\frac{\partial^3 f}{\partial \eta^3} + \frac{f}{2}\frac{\partial^2 f}{\partial \eta^2}\right) = 0
\end{equation}
Suppose that the solution of equation (28) is a series given by
\begin{equation}   
    f = F_0 + pF_1 + p^2F_2 + \ldots
\end{equation}

Substituting (29) into (28), we get
\begin{center}
    $\frac{\partial^3F_0}{\partial \eta^3} + p\frac{\partial^3 F_1}{\partial \eta^3} + p^2\frac{\partial^3 F_2}{\partial \eta^3} - \frac{\partial^3 f_0}{\partial \eta^3} + p\frac{\partial^3 f_0}{\partial \eta^3} + p\left[\frac{F_0}{2}\left(\frac{\partial^2 F_0}{\partial \eta^2} + p\frac{\partial^2 F_1}{\partial \eta^2}\right) + p\frac{F_1}{2}\left(\frac{\partial^2 F_0}{\partial \eta^2} + p\frac{\partial^2 F_1}{\partial \eta^2}\right)\right] + \ldots = 0$
\end{center}
Rearranging the coefficients of the terms with identical powers of \(p\), we have
\begin{align}
    p^0 &: \frac{\partial^3F_0}{\partial \eta^3} - \frac{\partial^3 f_0}{\partial \eta^3} = 0 \nonumber \\
    p^1 &: \frac{\partial^3 F_1}{\partial \eta^3} + \frac{\partial^3 f_0}{\partial \eta^3} + \frac{F_0}{2}\frac{\partial^2 F_0}{\partial \eta^2} = 0 \nonumber \\
    p^2 &: \frac{\partial^3 F_2}{\partial \eta^3} + \frac{F_1}{2}\frac{\partial^3 F_0}{\partial \eta^3} + \frac{F_0}{2}\frac{\partial^2 F_1}{\partial \eta^2} = 0 \nonumber \\
    p^3 &: \frac{\partial^3 F_3}{\partial \eta^3} + \frac{F_1}{2}\frac{\partial^3 F_1}{\partial \eta^3} + \frac{F_2}{2}\frac{\partial^2 F_0}{\partial \eta^2} + \frac{F_0}{2}\frac{\partial^2 F_2}{\partial \eta^2} = 0 \label{eq:p3}\\
    &\vdots \nonumber
\end{align}
First, we take \(F_0 = f_0\), and we start iterating by defining \(f_0\) as a Taylor series of order two at \(\eta = 0\) to make it accurate near \(\eta = 0\)
\begin{center}
    $F_0 = f_0 \frac{f''(0)}{2}\eta^2 + f'(0)\eta = f(0)$
\end{center}
Let us take \(f''(0) = 0.332057\), and from the given boundary conditions \(f(0) = 0\) and \(f'(0) = 0\). So,
\begin{center}
    $f_0 = \frac{0.332057}{2}\eta^2$
\end{center}
\begin{center}
    $f_0 = 0.1660285\eta^2$
\end{center}
Now, using this value to solve for \(F_1\) from (30):
\begin{center}
    $\frac{\partial^3 F_1}{\partial \eta^3} + \frac{\partial^3 f_0}{\partial \eta^3} + \frac{F_0}{2}\frac{\partial^2 F_0}{\partial \eta^2} = 0$
\end{center}
\begin{center}
    $\frac{\partial^3 F_1}{\partial \eta^3} = - \frac{F_0}{2}\frac{\partial^2 F_0}{\partial \eta^2}$
\end{center}
\begin{center}
    $= - \frac{0.1660285}{2}\eta \frac{\partial^2 (0.1660285)}{\partial \eta^2} \eta^2$
\end{center}
\begin{center}
    $\frac{\partial^3 F_1}{\partial \eta^3} = -(0.1660285)^2 \cdot \eta^2$
\end{center}
\begin{center}
    $F_1 = -(0.1660285)^2 \cdot \frac{\eta^5}{345}$
\end{center}
\begin{center}
    $\Longrightarrow F_1 = f_1 = -0.00045942 \cdot \eta^5$
\end{center}
Similarly, from (30), we can easily calculate the values of \(f_2, f_3, \ldots\):
\begin{center}
    $f_2 = 0.00000249 \cdot \eta^8$
\end{center}
\begin{center}
    $f_3 = 0.00000001 \cdot \eta^{11}$
\end{center}
For the assumption \(p=1\), we get
\begin{center}
    $f(\eta) = 0.1660285\eta^2 - 0.00045942\eta^5 + 0.00000249\eta^8 - 0.00000001\eta^{11}$ 
\end{center}
\subsection{Solution of Burger's Equation by Homotopy Perturbation Method}

To illustrate the modification algorithm of the HPM, consider the following nonlinear partial differential equation with a time derivative of any order:

\begin{multline}
D_t^n u(x,t) = L(u,u_x,u_{xx}) \\
+ N(u,u_x,u_{xx}) + f(x,t),  t > 0
\end{multline}

Where \(L\) is a linear operator, \(N\) is a nonlinear operator, and \(f\) is a known analytic function, subject to the initial conditions:

\begin{equation}
\frac{\partial^n}{\partial t^m}u(x,0) = h_m(x), \;\; m = 0,1,2,3,...n-1
\end{equation}

In view of the homotopy technique, we can construct the following homotopy:

\begin{multline}
\frac{\partial^n u}{\partial t^n} - L(u,u_x,u_{xx}) - f(x,t) \\
= p\left[\frac{\partial^n u}{\partial t^n} + N(u,u_x,u_{xx}) - D_t^n u\right]
\end{multline}

or

\begin{multline}
\frac{\partial^n u}{\partial t^n} - f(x,t) \\
= p\left[\frac{\partial^n u}{\partial t^n} + L(u,u_x,u_{xx}) + N(u,u_x,u_{xx}) - D_t^n\right]
\end{multline}

Where \(p\in[0,1]\). The homotopy parameter \(p\) always changes from zero to unity. When \(p=0\), (33) becomes the linearized equation:

\begin{equation}
\frac{\partial^n u}{\partial t^n}
 = L(u,u_x,u_{xx})+f(x,t)
\end{equation}
and (34) becomes the linearized equation:

\begin{equation}  
\frac{\partial^n u}{\partial t^n} = f(x,t)
\end{equation}

and when \(p = 1\), (33) and (34) turns out to be the original differential (31). The basic assumption is that the solution of (33) or (34) can be written as power series in p:
\begin{equation}
    u = u_0+pu_1+p^2u_2+...
\end{equation}

Finally, we approximate the solutions \(x,t)\) by:
\begin{equation}
    u(x,t) = \sum_{i=0}^{\infty}u_i(x,t)
\end{equation}

Consider the following one dimensional coupled Burger's equation
\begin{equation}
    u_t-U_{x,x}-uu_x+(uv)_x=0,
\end{equation}
\begin{equation}
    v_t-V{x,x}-2v{x,x}-2vv_x+(uv)_x=0
\end{equation}
With the initial conditions:
\begin{equation}
    u(x,0) = cosx, \;\;v(x,0) = cosx
\end{equation}
Making use of (34), the homotopy for (39) and (40) are:
\begin{equation}
    \frac{\partial u}{\partial t}= p\Bigg[\frac{\partial u}{\partial t}+u_{xx}+2uu_x-(uv)_x+D^u_t\Bigg]
\end{equation}
\begin{equation}
    \frac{\partial v}{\partial t}= p\Bigg[\frac{\partial v}{\partial t}+v_{xx}+2vv_x-(uv)_x+D^v_t\Bigg]
\end{equation}

As above, the basic assumption is that solution of (39) and (40) can be written as power series in\(p\)
\begin{equation}
    u = u_0+pu_1+p^2u_2+...
\end{equation}
\begin{equation}
    v = v_0+pv_1+p^2v_2+...
\end{equation}
Therefore substituting (44) and (45) and the initial condition (41) into (42) and (43) respectively and equating the terms with identical powers of \(p\), we can obtain the following set of linear partial differential equations.
\begin{center}
   $\frac{\partial u_0}{\partial t} = 0, \;\; u_0(x,0) = cosx$
\end{center}
\begin{center}
   $\frac{\partial v_0}{\partial t} = 0, \;\; v_0(x,0) = cosx$
\end{center}
\begin{center}
   $\frac{\partial u_1}{\partial t} =  \frac{\partial u_0}{\partial t} + (u_0)_{xx}-2u_0(u_0)_x-u_0(v_0)_x-v_0(u_0)_x-   D_tu_0, \;\; u_1(x,0) = 0$
\end{center}
\begin{center}
   $\frac{\partial v_1}{\partial t} =  \frac{\partial v_0}{\partial t} + (v_0)_{xx}-2v_0(u_0)_x-v_0(u_0)_x-D_tv_0, \;\; v_1(x,0) = 0$
\end{center}
\begin{center}
   $\frac{\partial u_2}{\partial t} =  \frac{\partial u_1}{\partial t} + (u_1)_{xx}+2u_0(u_1)_x+2u_1(u_0)_x-u_0(v_1)_x-u_1(v_0)_x-v_1(u_0)_x-v_0(u_1)_x-D_tu_1, \;\; u_2(x,0) = 0$
\end{center}
\begin{center}
   $\frac{\partial v_2}{\partial t} =  \frac{\partial v_1}{\partial t} + (v_1)_{xx}+2v_0(v_1)_x+2v_1(v_0)_x-u_0(v_1)_x-u_1(v_0)_x-v_1(u_0)_x-v_0(u_1)_x-D_tv_1, \;\; v_2(x,0) = 0$
\end{center}
\begin{center}
   $\frac{\partial u_3}{\partial t} =  \frac{\partial u_2}{\partial t} + (u_2)_{xx}+2u_0(u_2)_x+2u_1(u_1)_x+2u_2(u_0)_x-u_0(v_2)_x-u_1(v_1)_x-u_2(v_0)_x-u_0(v_2)_x-v_1(u_1)_x-v_2(u_0)_x-D_tu_2, \;\; u_3(x,0) = 0$
\end{center}
\begin{center}
$\frac{\partial v_3}{\partial t} = \frac{\partial v_2}{\partial t} + (v_2)_{xx} + 2v_0(v_2)_x + 2v_1(v_1)_x + 2v_2(v_0)_x - u_0(v_2)_x - u_1(v_1)_x - u_2(v_0)_x - v_0(u_2)_x - v_1(u_1)_x - v_2(u_0)_x - D_t v_2, \;\; v_3(x,0) = 0$
\end{center}
And so on.
Consequently, the first few components of the homotopy perturbation solution for (39) and (40) are derived as follows:
\begin{center}
$U_0(x,t) = \cos x , $
\end{center}

\begin{center}
   $ V_0(x,t) = cosx ,$
\end{center}
\begin{center}
   $ U_1(x,t) = cosx\cdot t,$
\end{center}
\begin{center}
   $ V_1(x,t) = cosx\cdot t,$
\end{center}
\begin{center}
   $ U_2(x,t) = cosx\cdot \frac{t}{2},$
\end{center}
\begin{center}
   $ V_2(x,t) = cosx\cdot \frac{t}{2},$
\end{center}
\begin{center}
   $ U_3(x,t) = -cosx\cdot \frac{t^3}{6},$
\end{center}
\begin{center}
$V_3(x,t) = -\cos x \cdot \frac{t^3}{6},$
\end{center}
and so on. In the same manner, the rest of the components can be obtained. The \(n\)-term approximation for equations (39) and (40) is given by
\begin{multline}
U(x,t) = \sum_{i = 0}^{n-1} u_i(x,t) = \cos x \Bigg[1 - t + \frac{t^2}{2!}\\ - \frac{t^3}{3!} + \dots \Bigg]
\end{multline}

\begin{multline}
V(x,t) = \sum_{i = 0}^{n-1} v_i(x,t) = \cos x \Bigg[1 - t + \frac{t^2}{2!}\\ - \frac{t^3}{3!} + \dots \Bigg]
\end{multline}
In closed form, this gives the solution:
\begin{equation}
U(x,t) = \cos x e^{-t}
\end{equation}
\begin{equation}
V(x,t) = \cos x e^{-t}
\end{equation}
Which is the exact solution to the one-dimensional coupled Burger's equations (39) and (40).
\section{Literature Review}
The Homotopy Perturbation Method (HPM) has emerged as a powerful tool for solving complex differential equations. Combining perturbation theory and homotopy analysis, HPM offers an effective approach to tackle nonlinear problems. A series of influential references shed light on the application of HPM and its impact.

"Nahfey's Introduction to Perturbation Method Technique" provides a fundamental introduction to perturbation methods, laying the groundwork for understanding HPM principles and techniques.

Liao's "Homotopy Analysis Method in Non-Linear Differential Equations" explores the homotopy analysis method, which serves as a theoretical foundation for HPM.

"He's Homotopy Perturbation Technique" represents a pivotal milestone in HPM development, explaining how to break down complex problems into simpler ones. This work extends HPM's applicability to various nonlinear equations.

In "Application of Homotopy Perturbation Method to Linear and Non-Linear Schrödinger Equations," Mohhamed and Shahwar demonstrate HPM's real-world utility in quantum mechanics, particularly with Schrödinger equations.

Ganji, Babazadeh, Noori, Pirouz, and Janipor, in their research titled "Utilizing the Homotopy Perturbation Method to Address the Non-Linear Blasius Equation in Boundary Layer Flow over a Flat Plate," employ HPM to demonstrate its effectiveness in the field of fluid dynamics.

Hameda's "Homotopy Perturbation for System of Non-Linear Coupled Equations" explores HPM's application in handling coupled nonlinear systems, highlighting its adaptability to diverse problem domains.

These works collectively provide valuable insights into the Homotopy Perturbation Method, covering its theoretical foundations, practical applications, and its ability to solve a variety of nonlinear equations. The references mentioned are sources of inspiration for further research and contribute to the ongoing development of mathematical problem-solving techniques.
\section{Conclusion}
In this project, a clear and mathematically supported conclusion drawn from the results is that HPM demonstrates rapid convergence towards exact solutions. It is noteworthy that HPM proves to be an effective, simple, and highly accurate tool for handling and solving Blasius and Burger equations, as well as various other types of nonlinear equations in a unified manner. Moreover, HPM distinguishes itself from traditional perturbation methods, which often rely on small parameter assumptions that may lead to non-physical results in many cases. Furthermore, numerical methods tend to yield inaccurate results when the equation exhibits strong time-dependence. In contrast, He’s Homotopy Perturbation Method (HPM) completely overcomes these shortcomings, highlighting the convenience and effectiveness of homotopy in resolving such mathematical challenges.
\section{References}
1. Nahfey, A.H. (1981). \textit{Introduction to the Perturbation Technique}. Wiley, New York.

2. Liao, S. (2011). \textit{Homotopy Analysis Method in Non-Linear Differential Equations}. Springer, Shanghai, China.

3. He, J.H. (1999). \textit{Homotopy Perturbation Technique}. \textit{Computer Methods in Applied Mechanics and Engineering}, 178, 257-262.

4. Mohammad, M.M., and Shahwar, F.R. (2008). \textit{Application of Homotopy Perturbation Method to Linear and Non-Linear Schrodinger Equation}. 63a, 140-144.

5. Ganji, D.D., Babazadeh, H., Noori, F., Pirouz, M.M., Janipour, M. (2009). \textit{An Application of Homotopy Perturbation Method for Non-Linear Blasius Equation to Boundary Layer Flow over a Flat Plate}. \textit{International Journal of Non-Linear Science}, 7, 309-404.

6. Hameda, A.A. (2012). \textit{Homotopy Perturbation for System of Non-Linear Coupled Equations}. 6, no. 96, 4787-4800.

\end{document}